\documentclass[12pt,a4paper,final]{iopart}

\usepackage{iopams}  
\expandafter\let\csname equation*\endcsname\relax
\expandafter\let\csname endequation*\endcsname\relax
\usepackage{amsmath}
\usepackage{graphicx}

\DeclareMathOperator*{\SumInt}{%
\mathchoice%
  {\ooalign{$\displaystyle\sum$\cr\hidewidth$\displaystyle\int$\hidewidth\cr}}
  {\ooalign{\raisebox{.14\height}{\scalebox{.7}{$\textstyle\sum$}}\cr\hidewidth$\textstyle\int$\hidewidth\cr}}
  {\ooalign{\raisebox{.2\height}{\scalebox{.6}{$\scriptstyle\sum$}}\cr$\scriptstyle\int$\cr}}
  {\ooalign{\raisebox{.2\height}{\scalebox{.6}{$\scriptstyle\sum$}}\cr$\scriptstyle\int$\cr}}
}

\usepackage[breaklinks=true,colorlinks=true,linkcolor=blue,urlcolor=blue,citecolor=blue]{hyperref}

\begin{document}

\title[]{On polynomial solutions to Fokker-Planck and sinked density evolution equations}

\author{Mathew Zuparic}
\address{Defence Science and Technology Organisation (DSTO), ACT 2600, Australia}
\ead{mathew.zuparic@dsto.defence.gov.au}

\begin{abstract}
We analytically solve for the time dependent solutions of various density evolution models. With specific forms of the diffusion, drift and sink coefficients, the eigenfunctions can be expressed in terms of hypergeometric functions. We obtain the relevant discrete and continuous spectra for the eigenfunctions. With non-zero sink terms the discrete spectra eigenfunctions are generalisations of well known orthogonal polynomials: the so-called \textit{associated-Laguerre, Bessel, Fisher-Snedecor} and \textit{Romanovski functions}. We use a MacRobert's proof to obtain closed form expressions for the continuous normalisation of the Romanovski density function. Finally, we apply our results to obtain the analytical solutions associated with the Bertalanffy-Richards Langevin equation.
\end{abstract}

\vspace{2pc}

\section{Introduction}
\label{intro}
The Fokker-Planck equation has been the focus of many decades of study due to its relevance in physics, finance, probability and statistics \cite{RISKEN,Schuss10}. \cite{Wong2} provides a particularly early example examining analytically tractable solutions to the Fokker-Planck equation, with more contemporary examples provided by \cite{Suvak3,BRICS,Suvak1,Suvak4,Linetsky,Linetsky2}.

Essentially, this work focuses on time dependent densities ${\cal T}(x,t|y)$, $t \ge 0$, of a diffusion process described by range $x$ given that it started at position $y$, governed by,
\begin{eqnarray}
\frac{\partial}{\partial t}{\cal T}(x,t|y)= \left\{ \frac{\partial^2}{\partial x^2}s(x)-\frac{\partial}{\partial x}q(x)-r(x) \right\}{\cal T}(x,t|y),\label{begin} \\
{\cal T}(x,0|y) = \delta(x -y),\;\; s(x) >0. \nonumber
\end{eqnarray}
Eq.(\ref{begin}) is defined on some interval on $\mathbb{R}$ with endpoints $(e_1,e_2)$ where $-\infty \le e_1 < e_2 \le \infty$. The continuous functions $s(x), q(x)$ and $r(x)$ are referred to as the diffusion, drift and the sink coefficients respectively. 

The applications of Eq.(\ref{begin}) are wide ranging. For $r(x) =0$, Eq.(\ref{begin}) is commonly referred to as the Fokker-Planck equation and one can readily show that ${\cal T}(x,t|y)$ is conserved for $t \ge 0$. The corresponding solutions to the Fokker-Planck equation are the time dependent probability densities associated with the (It\={o}) stochastic Langevin process,
\begin{eqnarray*}
\dot{x}(t) = q(x(t)) + \sqrt{2 s(x(t))} \eta(t), \;\; x(0) = y,
\end{eqnarray*}
where $\eta(t)$ is a Gaussian white noise term with unit variance. The Langevin equation is ubiquitous in a wide range of applications, from its beginnings in Brownian motion (see Chap.1 of \cite{Schuss10}), to finance \cite{Linetsky}, biological processes \cite{GARCIA} and the synchronisation of networked oscillators \cite{STROGATZ}. 

For $r(x) \ne 0$, the corresponding continuity equation is sinked, thus we expect the quantity being measured in Eq.(\ref{begin}) to seep away with time. It is conceptually important to develop analytically tractable solutions to Eq.(\ref{begin}) as they are Green's functions, which are both inherently mathematically interesting, and highly applicable - for an account of their application in physics see Chap.7 of \cite{MORFES}. Following \cite{FORRESTER}, Green's/density functions appearing in this work also figure heavily in random matrix theory. We shall indicate some of these connections throughout this work. Additionally, many past results for the conserved Fokker-Planck equation (we offer \cite{ZUPKAL} as a typical example) rely simply on the steady state density to gain insights. Sinked densities allow no such avenue for inquiry as their solutions decay with time. We shall highlight this behaviour in the proceeding sections. 

The general strategy for solving Eq.(\ref{begin}) is as follows: we first obtain the \textit{weight function} $W(x)$, found  by solving the corresponding Pearson's equation,
\begin{eqnarray}
\left\{ \frac{d}{d x} s(x) -  q(x)\right\}  W(x) =0 \;\; \Rightarrow \;\; 
W(x) = \frac{\kappa}{ s(x)} \exp \left\{  \int^x_{x'} d\xi \frac{q(\xi)}{s(\xi)} \right\},
\label{station}
\end{eqnarray}
for constants $\kappa$ and $x'$. Given the form of the diffusion, drift and sink coefficients, we categorise the spectrum of Eq.(1) as either discrete or mixed discrete/continuous. This gives us the general form of ${\cal T}(x,t|y)$ as,
\begin{eqnarray}
{\cal T}(x,t|y) =W(x) \SumInt_{\lambda} e^{-\lambda t}\rho_{\lambda}  \vartheta_{\lambda}(x) \vartheta_{\lambda}(y) ,\label{SOLLL} 
\end{eqnarray}
where the sum will be an integral for the continuous parts of the spectrum. We then find the corresponding eigenvalues and eigenfunctions of the system through standard techniques \cite{STEGUN,Koekoek,DLMF}. The final step involves solving for the normalisation constants which satisfy the initial condition. For the discrete spectrum eigenfunctions we utilise the orthogonal polynomial relation (Chap.3 of \cite{Koekoek}),
\begin{eqnarray}
\rho_n = \frac{1}{\int^{e_2}_{e_1}dx W(x)\vartheta^2_n (x) },\label{ORTHO1}
\end{eqnarray}
and for the continuous spectrum eigenfunctions we employ a MacRobert's inverse integral transform of the form,
\begin{eqnarray}
\int^{e_2}_{e_1} dx W(x)  \vartheta (\nu,x) \int^{\infty}_{0}d\mu \rho(\mu) \vartheta (\mu,x) = \rho(\nu) \Lambda(\nu). \label{MACCA}
\end{eqnarray}
See \cite{Wimp} for an instance involving the Whittaker functions, and \cite{HUSSAIN} and Chap.14 of \cite{DAVIES} for examples involving Bessel/Hankel functions. These inverse integral transforms usually rely on some key results attributable to MacRobert \cite{MACROBERT} which we shall exploit when deriving the continuous normalisation for the Romanovski case.

Given the pervasive nature of the Fokker-Planck equation, Green's functions and orthogonal polynomials, most of the cases presented in this work have been fully solved in the literature \textit{without} the sink term. What is new in this work is we present the full time dependent solutions for the sinked variants ($r(x) \ne 0$), and apply these results to obtain new solutions to  the stochastic Bertalanffy-Richards (B-R) equation \cite{BERT,RICH}. For an introduction to the application of the B-R equation in population modelling and biological processes see \cite{GARCIA}. 

In the next section we detail the forms of the diffusion, drift and sink coefficients that lead to the orthogonal polynomial eigenfunctions considered in this work. In Sec.\ref{1DSL} we give necessary information about Sturm-Liouville (S-L) operators, the Hilbert spaces their eigenfunctions span and how the sink terms in the S-L operators form associated variations of the orthogonal polynomials/eigenfunctions. In Sec.\ref{classification} we detail how the form of the S-L operator determines the exact form of the spectra for the eigenfunctions, along with the corresponding solutions to Eq.(\ref{begin}). In Sec.\ref{APPLICI} we apply the results of this work to the stochastic B-R equation. Finally we offer implications of these results and flag future work.

\section{Orthogonal polynomials}
\label{ORTHOG}
\subsection{Negative eigenvalues}
Applying the weight function, we decompose ${\cal T}(x,t|y)$ in Eq.(\ref{begin}) as,
\begin{eqnarray}
{\cal T}(x,t|y) =W(x) g(x,t|y). \label{DECOMP}
\end{eqnarray}
Hence the ensuing equation for $g(x,t|y)$ is
\begin{eqnarray*}
\frac{\partial}{\partial t}g(x,t|y) = {\cal H} g(x,t|y),
\end{eqnarray*}
where ${\cal H}$ is the S-L operator
\begin{eqnarray}
{\cal H} =  s(x)\frac{d^2}{d x^2} + q(x) \frac{d}{d x} -r(x).
\label{Sturm-Liouville}
\end{eqnarray}
We require that the operator ${\cal H}$ be negative, i.e. all relevant eigenfunctions of ${\cal H}$ have negative eigenvalues,
\begin{eqnarray}
{\cal H} \vartheta_{\lambda}(x) =- \lambda \vartheta_{\lambda}(x),\;\; \lambda \ge 0.
\label{S-Lspectrum}
\end{eqnarray}
The eigenvalues $\lambda$ may be discrete or continuous depending on boundary conditions \cite{Linetsky}, to be
specified explicitly in Sec.\ref{classification}. 

\subsection{Continuous classical orthogonal polynomials}
For this work $s(x), q(x)$ and $r(x)$ in Eq.(\ref{S-Lspectrum}) have the specific forms:
\begin{eqnarray}
s(x) \;\; \textrm{is at most quadratic in $x$},\nonumber\\
q(x) \;\; \textrm{is at most linear in $x$},\nonumber\\
r(x) \;\; \textrm{is either $0$, or $\lim_{x \rightarrow \{0,\infty\}}x r(x) = constant$}.\label{SPECIFICATIONS}
\end{eqnarray}
Focusing on the $r(x)=0$ case, Theorem 4 of \cite{Koekoek} states that there are six classes of discrete spectrum eigenfunctions to Eq.(\ref{S-Lspectrum}): Hermite, Laguerre, Jacobi, Bessel, Fisher-Snedecor (shifted-Jacobi), and Romanovski (pseudo-Jacobi) polynomials - the so-called \textit{continuous classical orthogonal polynomials}.  For $r(x)\ne 0$ the eigenfunctions generally have the same polynomial form, but are multiplied by the diffusion coefficient $s(x)$ raised to some power, $\varkappa$, specified in Sec.(\ref{1DSL}) of this work. We do not formally consider the Hermite and Jacobi case in this work (we only refer to them) as the Hermite case offers nothing new, and the Jacobi case only has finite support in $x$. In \cite{SAAD}, hypergeometric polynomial solutions to Eq.(\ref{S-Lspectrum}) for quite general forms of the diffusion and drift coefficients were studied. 

Using the Liouville transformation (Eq.(\ref{changevars}) of this work) to transform Eq.(\ref{S-Lspectrum}) into a corresponding Schr\"{o}dinger equation, we note that the Laguerre, Bessel and Romanovski potentials correspond to the \textit{Coulomb} (Chap.6 of \cite{MERZ}), \textit{Morse} \cite{MORSE} and \textit{trigonometric Scarf} \cite{SCARF} potentials, albeit with an additional parameter corresponding to the sink coefficient. 

More recently the Laguerre polynomials were applied as solutions to the non-linear Madelung fluid equation \cite{MADELUNG} and a Burger's equation with a time-dependent forcing term \cite{YADAV2}. Both \cite{COHERENT1} and \cite{CALO1} highlight the connection between the Laguerre polynomials and the algebra $su(1,1)$, which is then exploited to construct the coherent Laguerre function, and explore squeezed states in the Calogero-Sutherland model. \cite{Fakhri1} also highlights connections between Lie algebras and the associated Laguerre functions. Additionally, the non-sinked variant of Eq.(\ref{SOLLL}) for the Laguerre case features in the financial Cox-Ingersoll-Ross (CIR) model \cite{CIR1,CIR2}, amongst other applications.

With regards to the Bessel polynomials, in \cite{Fakhri2} the ladder operators of the associated Bessel functions were explored. The non-sinked variant of Eq.(\ref{SOLLL}) for the Bessel case was derived as the Fokker-Planck equation of an ergodic diffusion with reciprocal gamma invariant distribution in \cite{Suvak1}, and features in many financial models (see Sec.6.5 of \cite{Linetsky}). 

Following Chap.4 of \cite{Koekoek}, the Fisher-Snedecor polynomials are a variant of the Jacobi polynomials under a simple linear transformation such that the corresponding weight function's support is extended to the positive real line. In \cite{Suvak3} the non-sinked  variant of Eq.(\ref{SOLLL}) for the Fisher-Snedecor case was derived as the Fokker-Planck equation for an ergodic diffusion with Fisher-Snedecor invariant distribution. Refer to \cite{Wong2} for the corresponding Jacobi expression of Eq.(\ref{SOLLL}) which has finite support in $x$ and an entirely discrete spectrum.

The Romanovski polynomials have received a fair amount of attention lately due to their application in supersymmetric quantum mechanics \cite{QUESNE}, quantum chromodynamics \cite{Kirchbach2,Kirchbach1}, and connections with Yang-Mills integrals \cite{TIERZ}. The non-sinked variant of Eq.(\ref{SOLLL}) for the Romanovski case (without a closed form expression for the continuous spectrum normalisation) was first given in \cite{Suvak4} as the Fokker-Planck equation for an ergodic diffusion with the symmetric scaled Student invariant distribution.

Multidimensional generalisations of the classical polynomials of course exist (see \cite{KOORWINDER}, amongst other works) and are a current active field of study. Of particular relevance to this work we see in Chapters 2 and 3 of \cite{FORRESTER} that the probability density functions of the eigenvalues of the \textit{chiral}, \textit{Laguerre}, \textit{Jacobi} and \textit{Cauchy} ensembles of random matrices give the multidimensional generalisations of the Hermite, Laguerre, Jacobi and Romanovski weights respectively. Additionally, Chap.11 of the aforementioned work considers potentials which correspond to various classes of quantum Calogero-Sutherland models. In particular we see that Propositions 11.3.1 and 11.3.2 give multidimensional generalisations of the corresponding Schr\"{o}dinger Hermite, Laguerre and Jacobi potentials (amongst other more general cases), with the (restricted) Green's functions of these three cases constructed in Chap.11.6.

\section{One-dimensional Sturm Liouville operators}
\label{1DSL}
\subsection{Hilbert Space and finite orthogonality}
Due to ${\cal H}$ in Eq.(\ref{S-Lspectrum}) being a non-positive, self-adjoint S-L operator, the full set of solutions to Eq.(\ref{S-Lspectrum}) - present in Eq.(\ref{SOLLL}) - necessarily form a (weighted) square-integrable Hilbert space $L^2((e_2,e_1),W(x))$ with respect to the weighted inner product \cite{LANGER,McKEAN},
\begin{eqnarray}
\langle \vartheta_{\lambda}(x) | \vartheta_{\mu}(x) \rangle \equiv \int^{e_2}_{e_1}dx W(x) \vartheta_{\lambda}(x) \vartheta_{\mu}(x)< \infty. \label{FINITE1}
\end{eqnarray}
The emergence of the continuous spectrum in Eq.(\ref{SOLLL}) for certain classes of eigenfunctions is due to the discrete spectrum eigenfunctions possessing so-called \textit{finite orthogonality} (see Chapters 3 and 4 of \cite{Koekoek}): only a finite subset of the Bessel, Fisher-Snedecor and Romanovski polynomials obey Eq.(\ref{FINITE1}). Thus the continuous spectrum eigenfunctions are required to construct the Hilbert space. Following Chap.22 of \cite{HILPHIL} and Chapters 7 and 8 of \cite{REED}, if the spectrum of ${\cal H}$ is mixed, the corresponding Hilbert space is separable into the following orthogonal subspaces,
\begin{eqnarray}
L^2_{pp}((e_2,e_1),W(x)) \oplus L^2_{ac}((e_2,e_1),W(x)),\label{HILBERT}
\end{eqnarray}
where $L^2_{pp}$ denotes the subspace of the Hilbert space containing \textit{pure point} (discrete) spectrum, and $L^2_{ac}$ denotes the subspace of the Hilbert space containing \textit{absolutely continuous} spectrum.

\subsection{Associated orthogonal functions}
Recently in \cite{Fakhri1,Fakhri2,Fakhri3} the associated variants of the Laguerre, Bessel and Romanovski polynomials, respectively, were considered (the associated Fisher-Snedecor functions are a simple variation on the Romanovski case). The new results in this paper involve applying the aforementioned results and constructing the associated sinked densities. We list the canonical forms of the diffusion, drift and sink coefficients of the four relevant cases in Tab.\ref{TAB1}, and give the weight functions and the corresponding support of $x$ for each case in Tab.\ref{TAB2}.
\begin{table}[ht]
\caption{Canonical forms of $s(x)$, $q(x)$ and $r(x)$ considered in this work.} 
\centering 
\begin{tabular}{c c c c} 
\hline 
Case & $s(x)$ & $q(x)$ & $r(x)$ \\ [0.5ex] 
\hline 
Laguerre (L) & $x$ & $\sigma+1-x$ & $\frac{\gamma(\gamma+\sigma)}{x}$ \\
Bessel (B) & $x^2$ & $(\sigma+2)x+1$ & $\frac{\gamma}{x}$ \\
Fisher-Snedecor (F-S) & $x^2+x$ & $2(\sigma_1+1)x+\sigma_1+\sigma_2+1$ & $\frac{(\gamma-\sigma_1)(\gamma+\sigma_1+2\sigma_2(1+2x))}{4x(x+1)}$  \\
Romanovski (R) & $x^2+1$ & $2(\sigma_1+1)x+\sigma_2$ & $\frac{(\sigma_1-\gamma)(\gamma+\sigma_1-\sigma_2 x) }{x^2+1}$ \\ [1ex] 
\hline 
\end{tabular}
\label{TAB1} 
\end{table}
\begin{table}[ht]
\caption{Weight function $W(x)$ and corresponding support of $x$.} 
\centering 
\begin{tabular}{c c c } 
\hline 
Case & $W(x)$ & support \\ [0.5ex] 
\hline 
L & $x^{\sigma}e^{-x}$ & $x \in \mathbb{R}_+$  \\
B & $ x^{\sigma}e^{-\frac{1}{x}}$ & $x \in \mathbb{R}_+$  \\
F-S & $ x^{\sigma_1+\sigma_2}(x+1)^{\sigma_1-\sigma_2}$ & $x \in \mathbb{R}_+$   \\
R & $ (x^2+1)^{\sigma_1}e^{\sigma_2 \arctan(x)}$ & $x \in \mathbb{R}$  \\ [1ex] 
\hline 
\end{tabular}
\label{TAB2} 
\end{table}

We note that the inclusion of the sink expression adds a new parameter, $\gamma$, to each density equation. Past studies \cite{BOOK1,  BOOK2} have assured the negativity of the S-L operator by assuming $s(x)>0$ and $r(x) \ge 0$ for $x \in (e_1,e_2)$. In this work, due to the particular forms of $r(x)$ not following this restriction, we rely on the equivalent requirement: decomposing the eigenfunction,
\begin{eqnarray*}
\vartheta_{\lambda}(x) = s^{\varkappa}(x) \varphi_{\lambda}(x), \;\; \varkappa \in \mathbb{R},
\end{eqnarray*}
the corresponding S-L operator for $\varphi_{\lambda}(x)$ becomes,
\begin{eqnarray}
\bar{{\cal H}} \varphi_{\lambda}(x) \equiv  \left\{s(x)\frac{d^2}{d x^2} + \bar{q}(x) \frac{d}{d x}- \bar{r} \right\} \varphi_{\lambda}(x) =  - \lambda \varphi_{\lambda}(x),
 \label{S-Lspectrum2}
\end{eqnarray}
where
\begin{eqnarray}
 \bar{q}(x) = 2 \varkappa s'(x)+q(x),\nonumber\\
 \bar{r} = r(x)- \varkappa \left\{ s''(x) +\frac{(\varkappa-1)(s'(x))^2 + q(x) s'(x)}{s(x)} \right\}. \label{hatqr}
\end{eqnarray}
We require that $\bar{q}(x)$ is a also a linear function in $x$ and $\bar{r}$ is a positive constant. This guarantees that the original S-L operator ${\cal H}$ in Eq.(\ref{Sturm-Liouville}) is negative. In Tab.\ref{TAB3} we give $\varkappa$, and the ensuing expressions of $ \bar{q}(x)$ and $ \bar{r}$ for each case. We reiterate that in order for ${\cal H}$ to be negative, each $\bar{r}$ given in Tab.\ref{TAB3} needs to be positive. 
\begin{table}[ht]
\caption{Decomposition of the eigenfunction.} 
\centering 
\begin{tabular}{c c c c} 
\hline 
Case & $\varkappa$ & $ \bar{q}(x)$ & $\bar{r}$ \\ [0.5ex] 
\hline 
L & $\gamma$ & $2\gamma+\sigma+1-x$ & $\gamma$ \\
B & $ \frac{\gamma}{2}$ & $(2\gamma+\sigma+2)x+1$ &$-\gamma(\gamma+\sigma+1)$ \\
F-S & $\frac{\gamma-\sigma_1}{2}$ & $2(\gamma+1)x +\gamma+\sigma_2+1$  & $(\sigma_1-\gamma)(\gamma+\sigma_1+1)$ \\
R & $\frac{\gamma-\sigma_1}{2}$ & $2(\gamma+1)x +\sigma_2$ & $(\sigma_1-\gamma)(\gamma+\sigma_1+1)$  \\ [1ex] 
\hline 
\end{tabular}
\label{TAB3} 
\end{table}

The solutions to Eq.(\ref{S-Lspectrum2}) in the form of hypergeometric functions are standard in the mathematical literature. For many technical aspects of the details in the proceeding sections of this work we refer to \cite{Koekoek} for discrete spectrum eigenfunctions and \cite{STEGUN,DLMF} for the corresponding continuous spectrum eigenfunctions. We shall consider the eigen-spectra for each case explicitly in Sec.\ref{classification}.

\section{Spectral categories and solutions}
\label{classification}
The eigenvalue spectrum $\lambda$ of the S-L operator given in Eq.(\ref{Sturm-Liouville}) is determined by the behaviour of the operator at the boundaries. Two types of behaviour of $\cal{H}$ at the boundaries are relevant - designated \textit{non-oscillatory} and \textit{oscillatory}.

For each of the four instances, the range of $x$ is given either by $\mathbb{R}_+$ (Laguerre, Bessel and Fisher-Snedecor) or $\mathbb{R}$ (Romanovski). Hence the four cases have three possible boundaries: $\{0,\pm \infty\}$. The three S-L operators ${\cal H}$ which have a boundary at $0$ are classed as non-oscillatory at that boundary and require no special treatment. In the \textit{Feller boundary classification scheme} (see Chap.1 of \cite{SALMINEN} for instance), the remaining boundaries at $\pm \infty$ are classed as \textit{natural boundaries}\footnote{Not to be confused with the term \textit{natural boundary} regarding the analytic continuation of functions (see Chap 14.3 of \cite{COMPLEX}), amongst other uses of the term.} and require closer examination.
\subsection{The Liouville transformation} \label{LTrans}
 Transforming the variable $x$ and the eigenfunction $\vartheta_{\lambda}(x)$ via the following,
\begin{eqnarray}
z(x) = \int^x_{x'} \frac{d\xi}{\sqrt{s(\xi)}}, \;\; \varsigma_{\lambda}(z) = \vartheta_{\lambda}(x(z)) \sqrt{W(x(z)) \sqrt{s(x(z))} },
\label{changevars}
\end{eqnarray}
for constant $x'$, the corresponding S-L equation for $ \varsigma_{\lambda}(z)$ becomes the Schr\"{o}dinger equation,
\begin{eqnarray}
\left\{ \frac{d^2}{dz^2}- {\cal V}(x(z)) \right\}\varsigma_{\lambda}(z) = - \lambda \varsigma_{\lambda}(z), \label{varsig}
\end{eqnarray}
with the potential,
\begin{eqnarray}
{\cal V}(x) &=&\left( \frac{\frac{d}{dx}\sqrt{s(x)}}{2} \right)^2 -
\frac{\sqrt{s(x)} \frac{d^2}{dx^2}\sqrt{s(x)}}{2} + \frac{q^2(x)}{4 s(x)} + 
\frac{\frac{d}{dx}q(x)}{2}
\nonumber \\
&&- \frac{q(x) \frac{d}{dx}\sqrt{s(x)}}{\sqrt{s(x)}}+r(x),
\label{potential}
\end{eqnarray}
which includes the sink term $r(x)$. In Tab.\ref{TAB4} we list the relevant expressions regarding the Liouville transformation for our four cases. 
\begin{table}[ht]
\caption{Liouville transformations.} 
\centering 
\begin{tabular}{c c c c} 
\hline 
Case & $x(z)$ & $ {\cal V}(x(z))$ & $\sqrt{W(x(z)) \sqrt{s(x(z))} }$ \\ [0.5ex] 
\hline 
L & $ \frac{z^2}{4}$ & $\frac{z^2}{16} +\frac{4 \gamma(\sigma+\gamma)+ \sigma^2 -\frac{1}{4}}{z^2}-\frac{\sigma+1}{2}$ & $ \left(\frac{z}{2}\right)^{\sigma+\frac{1}{2}}e^{-\frac{z^2}{8}}$ \\
B & $e^z$ & $ \frac{(\sigma +2\gamma)e^{-z}}{2} + \frac{e^{-2z}}{4}+\left(\frac{\sigma+1}{2}\right)^2 $ &$\exp \left( \frac{\sigma+1}{2}z-\frac{e^{-z}}{2}\right)$ \\
F-S & $ \textrm{sinh}^2 \left( \frac{z}{2} \right)$ & $\frac{\gamma^2+\sigma^2_2- \frac{1}{4} +2 \gamma \sigma_2 \textrm{cosh}(z)}{\textrm{sinh}^2(z)}  +\left(\sigma_1+\frac{1}{2}\right)^2$  & $\left(\frac{\textrm{sinh}(z)}{2}\right)^{\sigma_1+\frac{1}{2}}\textrm{tanh}^{\sigma_2}\left( \frac{z}{2} \right)$ \\
R & $ \textrm{sinh}\left( z \right)$ & $\frac{ \frac{1}{4}- \gamma^2+\frac{\sigma^2_2}{4} +\gamma \sigma_2 \textrm{sinh}(z)}{\textrm{cosh}^2(z)}  +\left(\sigma_1+\frac{1}{2}\right)^2$ & $\textrm{cosh}^{\sigma_1+\frac{1}{2}}(z) e^{\frac{\sigma_2}{2}\arctan \{\textrm{sinh}(z)\}}$  \\ [1ex] 
\hline 
\end{tabular}
\label{TAB4} 
\end{table}

Given the Liouville transformations listed in Tab.\ref{TAB4}, the classification of the spectrum of ${\cal H}$ can now be given.
\subsection{Spectral classification}
Following Theorems 1, 2 and 3 of \cite{Linetsky}, the spectral properties depend on the behaviours of the diffusion $s(x)$ and the potential ${\cal V}(x)$ through the following specifications:

{ \it Classification at natural boundaries}
\begin{itemize}
\item{If $\lim_{x \rightarrow \pm \infty} z(x) \ne \pm \infty$, then ${\cal H}$ 
is classed as non-oscillatory at that boundary.}
\item{If $\lim_{x \rightarrow \pm \infty} z(x) = \pm \infty$, and 
$\lim_{z \rightarrow \pm \infty} {\cal V}(z) = \infty$, then ${\cal H}$ is 
classed as non-oscillatory at that boundary.}
\item{If $\lim_{z \rightarrow + \infty} {\cal V}(z) = \Lambda_{+} < \infty$, and/or  $\lim_{z \rightarrow - \infty} {\cal V}(z) = \Lambda_{-} < \infty$, and 
$\lim_{x\rightarrow \pm \infty} z^2(x)({\cal V}(x)-\Lambda_{\pm}) > - \frac{1}{4}$, then ${\cal H}$ at the corresponding boundary $(\pm \infty)$ is 
classed as non-oscillatory for $\lambda \in [0,\Lambda_{\pm}]$ 
and oscillatory for $\lambda > \Lambda_{\pm}$.}
\end{itemize}
Hence we have three possible spectral categories, which are detailed below.

\subsection{Spectral category I}\label{SPEC1} If ${\cal H}$ at both boundaries exhibits no oscillatory behaviour, then 
the spectrum is purely discrete and Eq.(\ref{SOLLL}) is given by,
\begin{eqnarray}
{\cal T}(x,t|y)= W(x)\sum^{\infty}_{n=0} e^{-\lambda_n t}\rho_n 
\vartheta_n (x) \vartheta_n (y),
\label{genexp}
\end{eqnarray}
where the normalisation coefficients are given by Eq.(\ref{ORTHO1}). We notice from Tab.\ref{TAB4} that the associated Laguerre functions $(\vartheta_n (x) \equiv x^{\gamma}L^{(2\gamma+\sigma)}_n(x))$ fall under this category. The hypergeometric form of the Laguerre polynomials are given by,
\begin{eqnarray*}
L^{(2\gamma+\sigma)}_n(x) &=&  \frac{(2\gamma + \sigma)_n}{n!}  \,_1F_1\left( \left. \begin{array}{c}
 -n\\
2\gamma + \sigma +1 \end{array} \right| x \right).
\end{eqnarray*}
Following Chap.9.12 of \cite{Koekoek} the eigenvalues and normalisation constants are given by,
\begin{eqnarray*}
 \lambda_n =n+\gamma,\;\; \rho_n= \frac{n!}{\Gamma(n+\sigma+2\gamma+1)}, \;\;\gamma \ge 0, \;\;  \sigma+2 \gamma > -1.
\end{eqnarray*}
Hence the expression of the density for the Laguerre case is,
\begin{eqnarray}
{\cal T}(x,t|y) &=&x^{\gamma+\sigma}y^{\gamma} e^{-x-\gamma t}\sum^{\infty}_{n=0}\frac{ n! e^{-nt}L^{(\sigma+2\gamma)}_n(x)L^{(\sigma+2\gamma)}_n(y)}{\Gamma(n+ \sigma+2\gamma+1)} \nonumber\\
&=& \frac{\left( \frac{x}{y} \right)^{\frac{\sigma}{2}}\exp\left\{\frac{\sigma}{2}t -\frac{x+ye^{-t}}{1-e^{-t}} \right\}}{1-e^{-t}}I_{\sigma+2\gamma}\left( \frac{2\sqrt{xy e^{-t}}}{1-e^{-t}} \right),\label{LAGRES}
\end{eqnarray}
where $I_{\alpha}(x)$ is the modified Bessel function of the first kind of order $\alpha$ (Chap.10 of \cite{DLMF}). The product form of the density is obtained through the application of the Hille-Hardy formula (Chap.18 of \cite{DLMF}).


\subsection{Spectral category II}\label{SPEC2} If ${\cal H}$ exhibits oscillatory behaviour at only one of the boundaries for $\lambda > \Lambda$, then Eq.(\ref{SOLLL}) is given by,
\begin{eqnarray}
{\cal T}(x,t|y)= W(x) 
\left\{
\sum^{N}_{n=0}e^{-\lambda_n t} \rho_n \vartheta_n (x) \vartheta_n (y)\right. \nonumber\\
\left.+ \int^{\infty}_{0}d \mu e^{-(\Lambda+\mu^2)t} \rho(\mu) \vartheta (\mu,x) \vartheta (\mu,y) 
\right\}, \;\; \lambda_N < \Lambda.
\label{genexp2}
\end{eqnarray}
where (following Lemmas 41 and 42 of \cite{BOOK1}) the eigenfunction with the continuous eigenvalue, $\vartheta (\mu,x)$, is the non-trivial solution to Eq.(\ref{S-Lspectrum}) which is square-integrable with $W(x)$ and valid in the neighbourhood of the boundary (in this case $\infty$) for which ${\cal H}$ exhibits the oscillatory behaviour. For this particular category we note that \cite{HILPHIL,REED},
\begin{eqnarray*}
 \vartheta_n (x) \in L^2_{pp}((e_2,e_1),W(x)),\;\; n \in \{ 0,1,\dots,N \}, \\
 \vartheta (\mu,x) \in L^2_{ac}((e_2,e_1),W(x)),\;\; \mu > 0.
\end{eqnarray*}
Since each respective subspace of the Hilbert space is orthogonal to the other, we are assured that any weighted inner product of a discrete and continuous eigenfunction is zero,
\begin{eqnarray*}
\langle \vartheta_n (x) |  \vartheta (\mu,x) \rangle \equiv \int^{e_2}_{e_1}dx W(x) \vartheta_n (x)  \vartheta (\mu,x) =0.
\end{eqnarray*}
The discrete normalisation constants in Eq.(\ref{genexp2}) are given by Eq.(\ref{ORTHO1}) and the continuous normalisation $\rho(\mu)$ can be obtained through the application of the MacRobert's inverse integral transform in Eq.(\ref{MACCA}).

From the forms of ${\cal V}(z)$ in Tab.\ref{TAB4} and the corresponding support of $x$, we see that the Bessel and Fisher-Snedecor cases fall under this particular mixed spectral category. Additionally we see from Tab.\ref{TAB4} that the highest discrete eigenvalue for each case is constrained by,
\begin{eqnarray*}
\lambda_N < \left\{  \begin{array}{cl} \left( \frac{\sigma+1}{2} \right)^2 & \textrm{Bessel}\\
 \left(\sigma_1+ \frac{1}{2} \right)^2 & \textrm{Fisher-Snedecor}
\end{array}\right.
\end{eqnarray*}
Addressing the discrete spectrum eigenfunctions, the associated Bessel function $(\vartheta_n (x) \equiv x^{\gamma} B^{(2\gamma+\sigma)}_n(x))$ and associated Fisher-Snedecor functions $(\vartheta_n (x) \equiv (x^2+x)^{\frac{\gamma-\sigma_1}{2}}F^{(\gamma,\sigma_2)}_n(x))$ have the following hypergeometric forms,
\begin{eqnarray*}
  B^{(2 \gamma + \sigma)}_{n}(x) &=&  \,_2F_0\left( \left. \begin{array}{cc}
-n,& 2\gamma+\sigma+n+1 \end{array}  \right| -x \right), \\
 F^{(\gamma,\sigma_2)}_{n}(x) &=& ( \gamma+\sigma_2+1)_n \,_2F_1\left( \left. \begin{array}{ll}
-n,& 2\gamma+n+1\\
& \gamma+\sigma_2+1 \end{array}  \right| -x \right).
\end{eqnarray*}
Following \cite{Suvak3} and Chap.9.12 of \cite{Koekoek}, the eigenvalues and normalisation constants for each case is given in Tab.\ref{TAB6}.
\begin{table}[ht]
\caption{Discrete eigenvalues and normalisation for Bessel and Fisher-Snedecor.} 
\centering 
\begin{tabular}{c c c c} 
\hline 
Case & $\lambda_n$ & $ \rho_n$ & restrictions \\ [0.5ex] 
\hline 
B & $ -(\gamma+n)(\gamma+\sigma+n+1)$ & $\frac{(-2n -2\gamma- \sigma -1)}{n!\Gamma(-n-2\gamma-\sigma)}$ & $ \gamma(\gamma+\sigma+1)\le 0$\\
&&& $\sigma+2 \gamma < -1$\\
&&& $n < -\gamma-\frac{\sigma+1}{2}$ \\
F-S & $(\sigma_1-\gamma-n)(\gamma+\sigma_1+n+1)  $ & $\frac{(-2n-1 -2\gamma)\Gamma(-n-\gamma+\sigma_2)}{n!\Gamma(-n-2\gamma)\Gamma(1+n+\gamma+\sigma_2)}$ & $(\sigma_1-\gamma)(\gamma+\sigma_1+1)\ge 0$\\
&&&  $2\gamma<-1$ \\
&&& $n <-\gamma-\frac{1}{2}$\\ [1ex] 
\hline 
\end{tabular}
\label{TAB6} 
\end{table}

Addressing the continuous spectrum eigenfunctions, the Bessel $(\vartheta (\mu,x) \equiv x^{\gamma} \psi_B(\mu,x)$ (see \cite{Suvak1} and Chap.13 of \cite{STEGUN,DLMF}) and Fisher-Snedecor $(\vartheta (\mu,x) \equiv (x^2+x)^{\frac{\gamma-\sigma_1}{2}} \psi_F(\mu,x)$ (see \cite{Suvak3} and Chap.15 of \cite{STEGUN,DLMF}) cases have the following hypergeometric forms,
\begin{eqnarray*}
\psi_B(\mu,x) = \left\{ \begin{array}{ll} \,_2F_0\left(\left.\begin{array}{cc} \gamma+\frac{\sigma+1}{2}-i \mu,& \gamma+\frac{\sigma+1}{2}+i \mu \end{array} \right| -x  \right) & |x|\le 0\\
\frac{\Gamma(-2 i \mu)}{\Gamma \left(\gamma+ \frac{\sigma+1}{2}- i \mu \right)} 
\tilde{\psi}_B(\mu,x)+ \frac{\Gamma(2 i \mu)}{\Gamma \left(\gamma+ \frac{\sigma+1}{2}+ i \mu \right)} \tilde{\psi}_B(-\mu,x) & |x|>0 \end{array} \right.\\
\psi_F(\mu,x) = \left\{ \begin{array}{ll} \,_2F_1\left(\left. 
\begin{array}{cc} \gamma+\frac{1}{2}+ i \mu, &\gamma+\frac{1}{2} - i \mu\\ 
& \gamma+1 +\sigma_2 \end{array}  \right| -x \right) & |x|\le 1\\
 \Pi(\mu)\tilde{\psi}_F(\mu,x) +  \Pi(-\mu)\tilde{\psi}_F(-\mu,x) & |x|>1 \end{array} \right. 
\end{eqnarray*}
where,
\begin{eqnarray*}
\tilde{\psi}_B(\mu,x)  =\left( \frac{1}{x} \right)^{\gamma+\frac{\sigma+1}{2}+ i \mu}\,
_1F_1\left(\begin{array}{c} \gamma+\frac{\sigma+1}{2}+ i\mu \\ 1 + 2i\mu \end{array} \left| 
\frac{1}{x} \right. \right),\\
\tilde{\psi}_F(\mu,x)=  \left(\frac{1}{x}\right)^{\gamma+\frac{1}{2}+i\mu} \,_2F_1\left(\left. \begin{array}{ll}\gamma+\frac{1}{2}+ i \mu, &\frac{1}{2} - \sigma_2+ i \mu\\ 
& 1+2i\mu \end{array}  \right| -\frac{1}{x} \right),\\
\Pi(\mu) =  \frac{\Gamma(\gamma+1+\sigma_2)\Gamma(-2i\mu)}{\Gamma(\gamma + \frac{1}{2}-i \mu)\Gamma(\frac{1}{2}+\sigma_2-i\mu)},
\end{eqnarray*}
and the corresponding eigenvalues and normalisations (\cite{Suvak3,Suvak1,Wimp}) are given in Tab.\ref{TAB7}.
\begin{table}[ht]
\caption{Continuous eigenvalues and normalisation for Bessel and Fisher-Snedecor.} 
\centering 
\begin{tabular}{c c c c} 
\hline 
Case & $\lambda(\mu)$ & $\rho(\mu)$ & restrictions  \\ [0.5ex] 
\hline 
B & $\left( \frac{\sigma+1}{2} \right)^2+\mu^2$ & $ \frac{\left|\Gamma\left( \gamma+\frac{\sigma+1}{2}+i\mu \right)\right|^2  }{2 \pi  \left| \Gamma(2i\mu) \right|^2}$ & $\mu>0$ \\
F-S & $\left( \sigma_1+\frac{1}{2} \right)^2+\mu^2$ & $ \frac{1}{2 \pi \left| \Pi(\mu) \right|^2}  $ & $\mu >0$  \\ [1ex] 
\hline 
\end{tabular}
\label{TAB7} 
\end{table}

Hence, the expression for the density of the Bessel case is,
\begin{eqnarray}
{\cal T}(x,t|y) =x^{\gamma+\sigma} y^{\gamma} e^{-\frac{1}{x}}\left\{ \sum^{\lfloor -\gamma- \frac{\sigma+1}{2}\rfloor}_{n=0}\frac{(-2\gamma -\sigma-2n-1)  }{n!\Gamma(-2\gamma - \sigma-2n)} \right. \nonumber \\
 \times e^{(\gamma+n)(\gamma+\sigma+n+1)t} B^{(2\gamma+\sigma)}_n(x) B^{(2\gamma+\sigma)}_n(y) \nonumber\\
\left. + \int^{\infty}_{0}d \mu \frac{\left| \Gamma \left(\gamma+\frac{\sigma+1}{2}+i\mu \right)\right|^2 e^{-\left\{\left( \frac{\sigma+1}{2} \right)^2+\mu^2  \right\}t}}{2 \pi\left| \Gamma\left( 2i\mu \right) \right|^2}   \psi_B(\mu,x)\psi_B(\mu,y) \right\}, \label{BESSRES}
\end{eqnarray}
and the corresponding expression for the Fisher-Snedecor case is,
\begin{eqnarray}
{\cal T}(x,t|y) =x^{\frac{\gamma+\sigma_1}{2}+\sigma_2} (x+1)^{\frac{\gamma+\sigma_1}{2}-\sigma_2} (y^2+y)^{\frac{\gamma-\sigma_1}{2}}\nonumber\\
\times\left\{ \sum^{\lfloor -\gamma- \frac{1}{2}\rfloor}_{n=0}\frac{(-2n-1 -2\gamma)\Gamma(-n-\gamma+\sigma_2)}{n!\Gamma(-n-2\gamma)\Gamma(1+n+\gamma+\sigma_2)}\right. e^{ (\gamma+n-\sigma_1)(\gamma+\sigma_1+n+1)t}   \nonumber \\
\left. \times F^{(\gamma,\sigma_2)}_n(x) F^{(\gamma,\sigma_2)}_n(y)  + \int^{\infty}_{0}d \mu \frac{e^{-\left\{\left( \sigma_1+\frac{1}{2} \right)^2+\mu^2  \right\}t}}{2 \pi\left| \Pi(\mu) \right|^2} \psi_F(\mu,x)\psi_F(\mu,y) \right\} . \label{FISHRES}
\end{eqnarray}

\subsection{Spectral category III}\label{SPEC3} If ${\cal H}$ exhibits oscillatory behaviour at both boundaries for $\lambda > \Lambda_{\pm}$, and $\Lambda_+=\Lambda_-$,
then  Eq.(\ref{SOLLL}) is given by,
\begin{eqnarray}
{\cal T}(x,t|y)=W(x) 
\left\{
\sum^{N}_{n=0}e^{-\lambda_n t} \rho_n \vartheta_n (x) \vartheta_n (y) \right.\nonumber\\
\left.+ \sum^2_{i,j=1}\int^{\infty}_{0}  d\mu e^{-(\Lambda + \mu^2) t} \rho_{i,j}(\mu) 
 \vartheta_i (\mu,x)  \vartheta_j (\mu,y)
\right\}, \;\; \lambda_N < \Lambda.
\label{genexp3}
\end{eqnarray}
where (following Sec.5.3 of \cite{Linetsky}) the eigenfunctions with the continuous eigenvalues, $\vartheta_i (\mu,x)$, $i=\{1,2\}$, are the linearly independent solutions to Eq.(\ref{S-Lspectrum}) which are square-integrable with $W(x)$, and are valid in the neighbourhood of the natural boundaries (in this case, $\pm \infty$) for which ${\cal H}$ exhibits oscillatory behaviour. Similar to spectral category II, we note that \cite{HILPHIL,REED},
\begin{eqnarray*}
 \vartheta_n (x) \in L^2_{pp}((e_2,e_1),W(x)),\;\; n \in \{ 0,1,\dots,N \}, \\
 \vartheta_i (\mu,x) \in L^2_{ac}((e_2,e_1),W(x)),\;\; i\in\{1,2\}, \;\; \mu > 0.
\end{eqnarray*}
The discrete normalisation constants in Eq.(\ref{genexp3}) are given by Eq.(\ref{ORTHO1}), and in \ref{ROMNORMAPP} we explicitly derive the continuous normalisations $\rho_{i,j}(\mu)$, using the aforementioned MacRobert's style proof. 

From the forms of ${\cal V}(z)$ in Tab.\ref{TAB4} and the corresponding support of $x$, we see that the Romanovski case fall under this particular mixed spectral category, and its highest discrete eigenvalue satisfies,
\begin{eqnarray*}
\lambda_N <  \left(\sigma_1+ \frac{1}{2} \right)^2.
\end{eqnarray*}
Concerning the discrete spectrum eigenfunctions, the associated Romanovski functions $(\vartheta_n (x) \equiv (x^2+1)^{\frac{\gamma-\sigma_1}{2}}R^{(\gamma,\sigma_2)}_n(x))$ have the following hypergeometric form,
\begin{eqnarray*}
 R^{(\gamma,\sigma_2)}_{n}(x)= (-2 i)^n \frac{\left(\gamma +i 
\frac{\sigma_2}{2}+1 \right)_n}{(n+2 \gamma +1)_n}\,_2F_1\left( \left. \begin{array}{ll}
-n,& 2\gamma+n+1\\
& \gamma+i \frac{\sigma_2}{2}+1 \end{array}  \right| \frac{1-ix}{2} \right).
\end{eqnarray*}
Following Chap.9.9 of \cite{Koekoek}, the eigenvalues and normalisation constants for this case are given by,
\begin{eqnarray}
\lambda_n &=&  (\sigma_1-\gamma-n)(\gamma+\sigma_1+n+1), \nonumber\\
\rho_n  &=&\frac{\Gamma(-n-2\gamma)\left|\Gamma\left(-\gamma-n+i\frac{\sigma_2}{2}\right)\right|^2}{2^{2(n+\gamma)+1}n!\Gamma(-2n-2\gamma-1)\Gamma(-2n-2\gamma)},
\label{ROMANDIS}
\end{eqnarray}
under the restrictions,
\begin{eqnarray*}
(\sigma_1-\gamma)(\gamma+\sigma_1+1) \ge 0, \;\; 2\gamma  < -1, \;\;n <-\gamma-\frac{1}{2}. 
\end{eqnarray*}
Rescaling the continuous spectrum eigenfunctions, 
\begin{eqnarray*}
\vartheta_1 (\mu,x) \equiv (x^2+1)^{\frac{\gamma-\sigma_1}{2}} \chi_1(\mu,x), \;\; \vartheta_2 (\mu,x) \equiv (x^2+1)^{\frac{\gamma-\sigma_1}{2}} \chi^*_1(\mu,x),
\end{eqnarray*}
(where $\chi^*$ is the complex conjugate of $\chi$), the eigenvalues are parameterised by,
\begin{eqnarray*}
\lambda(\mu)= \left(\sigma_1+ \frac{1}{2} \right)^2+\mu^2, \;\; \mu>0
\end{eqnarray*}
and hypergeometric forms of $\chi$ are (see \cite{Suvak4} and Chap.15 of \cite{STEGUN,DLMF}),
\begin{eqnarray}
 \chi_{1}(\mu,x) = \left\{ \begin{array}{ll} \,_2F_1\left(\left. 
\begin{array}{cc} \gamma+\frac{1}{2}+ i \mu, &\gamma+\frac{1}{2} - i \mu\\ 
& \gamma+1 +  i \frac{\sigma_2}{2} \end{array}  \right| \frac{1 - ix}{2} \right) & |x| \le \sqrt{3}\\
\tilde{\Gamma}(\mu) \tilde{\chi}(\mu,x) +\tilde{\Gamma}(-\mu) \tilde{\chi}(-\mu,x)& |x| > \sqrt{3} \end{array} \right. \label{CHI12}
\end{eqnarray}
where,
\begin{eqnarray}
\tilde{\chi}(\mu,x) =   \left(\frac{ix-1}{2} \right)^{-\gamma-\frac{1}{2}-i\mu} \,_2F_1\left(\left. \begin{array}{cc} \gamma+\frac{1}{2}+ i \mu, &\frac{1}{2} -i \frac{\sigma_2}{2}+ i \mu\\ 
& 1+2i\mu \end{array}  \right| \frac{2}{1-ix} \right),\nonumber\\
\tilde{\Gamma}(\mu) = \frac{\Gamma(\gamma+1+i\frac{\sigma_2}{2})\Gamma(-2i\mu)}{\Gamma(\gamma + \frac{1}{2}-i \mu)\Gamma(\frac{1}{2}+i \frac{\sigma_2}{2}-i\mu)}.\label{convenientgamma}
\end{eqnarray}
The continuous orthogonality relations are given by, 
\begin{eqnarray*}
\int^{\infty}_{-\infty} dx (x^2+1)^{\gamma}e^{\sigma_2 \arctan (x)} \chi_i(\nu,x) \int^{\infty}_{0}d\mu \rho_{i,j}(\mu) \chi_{j}(\mu,x) = \rho_{i,j}(\nu) \Lambda_{i,j}(\nu),
\end{eqnarray*}
where $\nu \in \mathbb{R}_+$ and,
\begin{eqnarray}
\Lambda_{1,1}(\mu) =\Lambda_{2,2}^*(\mu)= 2^{2\gamma+3} \pi \tilde{\Gamma}(\mu)\tilde{\Gamma}(-\mu) \cosh \left[\frac{\pi}{2} \left\{\sigma_2-i(2\gamma+1)  \right\} \right],\nonumber\\
\Lambda_{1,2}(\mu)=\Lambda_{2,1}(\mu) = 2^{2\gamma+2} \pi \left\{ \left| \tilde{\Gamma}(\mu) \right|^2 \cosh\left[ \frac{\pi}{2} (\sigma_2 + 2 \mu)  \right] \right.\nonumber\\
\left. +  \left| \tilde{\Gamma}(-\mu) \right|^2 \cosh\left[ \frac{\pi}{2} (\sigma_2 - 2 \mu)  \right]\right\}.\label{NEWRESULT}
\end{eqnarray}
Thus the continuous normalisations are given by,
\begin{eqnarray}
\rho_{1,1}(\mu) = \rho^*_{2,2}(\mu) = \frac{\Lambda_{2,2}(\mu)}{|\Lambda_{1,1}(\mu)|^2-\Lambda^2_{1,2}(\mu)},\nonumber \\
\rho_{1,2}(\mu) = \rho_{2,1}(\mu) = \frac{\Lambda_{1,2}(\mu)}{\Lambda^2_{1,2}(\mu)-|\Lambda_{1,1}(\mu)|^2}.\label{ROMRHO}
\end{eqnarray}
We provide the detailed derivation of Eq.(\ref{NEWRESULT}) in \ref{ROMNORMAPP} using the aforementioned MacRobert's method.

\begin{figure}
\begin{center}
\includegraphics[width=130mm,height=50mm,angle=0]{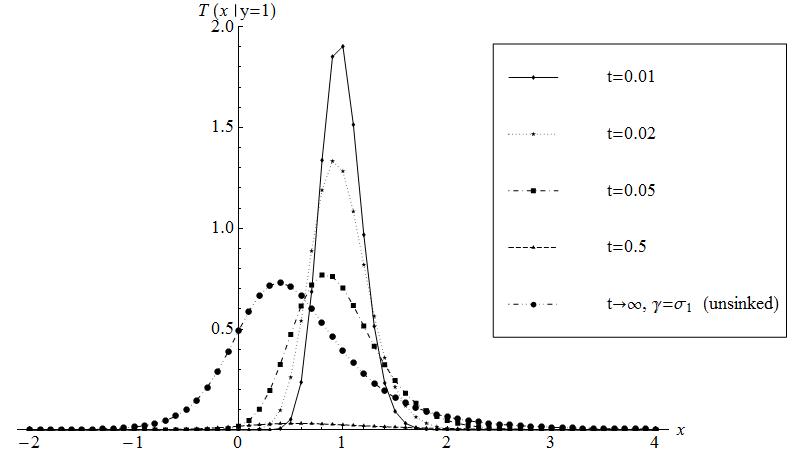}
\caption{Examples of Eq.(\ref{ROMRES}) at various times with parameter values $\sigma_1=-2.7$, $\sigma_2=2.1$ and $\gamma=-0.6$.} \label{fig2}
\end{center}
\end{figure}

Hence the complete density function for the Romanovski case is,
\begin{eqnarray}
{\cal T}(x,t|y) =(x^2+1)^{\frac{\gamma+\sigma_1}{2}}e^{\sigma_2 \arctan(x)}(y^2+1)^{\frac{\gamma-\sigma_1}{2}}\left\{ \sum^{\lfloor -\gamma- \frac{1}{2}\rfloor}_{n=0}\frac{\Gamma(-n-2\gamma)}{2^{2(n+\gamma+1)} \pi n! } \right. \nonumber\\
 \times \frac{\left|\Gamma\left(-n-\gamma+i\frac{\sigma_2}{2}\right)\right|^2 e^{ (\gamma+n-\sigma_1)(\gamma+\sigma_1+n+1)t}}{\Gamma(-2n-2 \gamma) \Gamma(-2n-2\gamma-1)}R^{(\gamma,\sigma_2)}_n(x)R^{(\gamma,\sigma_2)}_n(y) \nonumber \\
\left. + \sum^{2}_{k,l=1}\int^{\infty}_{0}d \mu \rho_{k,l}(\mu)e^{-\left\{\left(\sigma_1+ \frac{1}{2} \right)^2+\mu^2 \right\}t} \chi_k(\mu,x)\chi_l(\mu,y) \right\}.\label{ROMRES}
\end{eqnarray}
We give an example of Eq.(\ref{ROMRES}) in Fig.\ref{fig2}. Notice that as time increases the total area of the density (which begins at unity) decreases. For $t\ge 0.5$ we notice that the density is barely distinguishable visually. We compare this to the stationary density of the non-sinked case - the left most density - where area is conserved for all $t$.

\section{Application - Bertalanffy-Richards Langevin equation}
\label{APPLICI}
We now present an application of this work - time dependent distributions corresponding to various instances of the B-R Langevin equation. The B-R equation is a deterministic system given by,
\begin{eqnarray*}
\dot{x}(t) = a x (t) - b x^{\zeta} (t),\;\; \zeta > 1, \;\; x(0) = y,
\end{eqnarray*}
where $ \{ a,b\} \in \mathbb{R}_+ \cup 0$. We note that when $\{a,b\} > 0$, the $a x$ and $b x^{\zeta}$ terms act as growth and decay terms respectively; the greater the value of $\zeta$, the more pronounced the decay. The choice $\zeta = 2$ gives the famous logistic equation.  
\subsection{Stochastic perturbations and the Fokker-Planck equation}
To proceed we consider the two uncorrelated noise terms $\eta_1$ and $\eta_2$ with variance $\Omega$,
\begin{eqnarray*}
\langle\eta_i(t) \rangle =0,\;\; \langle\eta_i(t_1) \eta_j(t_2) \rangle =\delta_{ij} \Omega  \delta(t_1-t_2),
\end{eqnarray*}
where $\langle \dots \rangle$ means an ensemble average over the noise. We perturb the growth and decay coefficients by $\eta_1$ and $\eta_2$ respectively to obtain the following (It\={o}) stochastic Langevin equation, 
\begin{eqnarray}
\dot{x}(t) = \left\{a + \alpha \eta_1(t)\right\}x(t) - \left\{ b + \beta \eta_2(t) \right\} x^{\zeta}(t), \label{ITO}
\end{eqnarray}
where $ \{\alpha,\beta\} \in \mathbb{R}$. Eq.(\ref{ITO}) can be \textit{solved exactly} through the transformation
\begin{eqnarray}
x^{1-\zeta} =\xi, \;\;y^{1-\zeta} =\xi' , \label{ITOTRANS}
\end{eqnarray}
leading to the linear Langevin equation,
\begin{eqnarray*}
&&\dot{\xi}(t) = (\zeta-1)\left[b + \beta  \eta_2(t)  - \left\{a + \alpha  \eta_1(t)\right\}\xi(t)\right],
\end{eqnarray*}
and the formal solution,
\begin{eqnarray*}
\xi(t) = \frac{\left\{(\zeta-1)\int^t_0 d\tau_1  \{b+\beta  \eta_2(\tau_1)\} e^{(\zeta-1)\int^{\tau_1}_{0}d\tau_2 (a+ \alpha  \eta_1(\tau_2))}  +\xi'\right\}}{ e^{(\zeta-1)\int^{t}_{0}d\tau (a+\alpha  \eta_1(\tau)) }} .
\end{eqnarray*}
We refer to the above solution for $\xi(t)$ as formal as it contains integrals of specific instances of the noise terms (meaning that each solution will be different for different noise instances). In order to make general statements about the above system, we shall construct its probability density function. Following Chap.4.5 of \cite{Schuss10}, the stochastic process in Eq.(\ref{ITO}) obeys the following Fokker-Planck equation,
\begin{eqnarray}
\frac{\partial}{\partial t}{\cal T}(x,t|y) = \left\{  \frac{\partial^2}{\partial x^2}s(x) - \frac{\partial}{\partial x} q(x)\right\} {\cal T}(x,t|y),\;\; {\cal T}(x,0|y) = \delta(x-y)\nonumber,\\
\textrm{where}\;\;\;s(x) = \frac{\Omega}{2}\left(\alpha^2 x^2  + \beta^2 x^{2\zeta}\right) ,\;\; q(x) =  a x - b x^{\zeta}. \label{qands}
\end{eqnarray}
As mentioned in earlier sections, since the above equation contains no sink term ${\cal T}(x,t|y)$ is conserved, and its most natural interpretation is density of probability, where $x$ and $y$ are the population of a species.
\subsection{Restricting the Heun equation}
Our goal of this section is to analytically solve for various cases of Eq.(\ref{qands}) using our polynomial solutions for density functions given in Sec.\ref{classification}. Applying the standard decomposition in Eq.(\ref{DECOMP}), and the nonlinear transformation in Eq.(\ref{ITOTRANS}), the resulting expression for $g(\xi,t|\xi')$ is,
\begin{eqnarray}
\frac{\partial}{\partial t}g(\xi,t|\xi') =(\zeta-1) \left\{  \frac{\Omega(\zeta-1)}{2} \left( \alpha^2 \xi^2 + \beta^2 \right)\frac{\partial^2}{\partial \xi^2}\right.\nonumber\\
 \left.+\left(  \left[\frac{\Omega \zeta \alpha^2-2a}{2} \right] \xi +b + \frac{\Omega \zeta \beta^2}{2 \xi}  \right)\frac{\partial}{\partial \xi} \right\}g(\xi,t|\xi').\label{HUEN}
\end{eqnarray}
Since the B-R equation is used extensively in population modelling, where the variable $x$ represents the number of living members of a species, only eigenfunctions in the range $\mathbb{R}_+$ will be considered, hence leaving out the Romanovski example. This leaves three relevant cases, Laguerre, Bessel and Fisher-Snedecor.

The S-L operator on the right hand side of Eq.(\ref{HUEN}) leads to the Heun differential equation (see Chap.31 of \cite{DLMF}). Due to the Heun equation possessing four distinct singular points, there is no equivalent hypergeometric closed form expression for the Heun functions \cite{HEUN1}. Nevertheless, we find the following mapping between the Heun system and hypergeometric solutions:
\begin{itemize}
\item{$\alpha=b=0$ leads to the Laguerre case}
\item{$\beta=0$ leads to the Bessel case}
\item{$b=0$ leads to the Fisher-Snedecor case}
\end{itemize}
We shall only detail the Laguerre and Fisher-Snedecor cases in this work as the Bessel case was first solved in \cite{LAST2} and along with \cite{Wong2} is one of the earlier results involving analytical expressions of densities with mixed spectra. The case $\alpha=0$ in Eq.(\ref{HUEN}) leads to the Biconfluent Heun equation, whose solution suffers the same non-closed properties as the Heun equation (see Chap.31 of \cite{DLMF}). Additionally, the case $a=\alpha=0$ leads to the \textit{Bessel process with constant drift} \cite{Linetsky2}, which is a peculiar hypergeometric case (beyond the scope of this work) where the spectrum is mixed but the discrete part contains an \textit{infinite} number of eigenvalues.
\subsection{Laguerre case}
Setting $\alpha=b=0$, the weight function $W(\xi)$ for this case is,
\begin{eqnarray*}
W(\xi)=\xi^{\frac{2 \zeta}{\zeta-1}} e^{ - \frac{\xi^2}{\omega_L} },\;\; \omega_L=\frac{\Omega (\zeta-1) \beta^2}{a} .
\end{eqnarray*}
Applying the following change in variables,
\begin{eqnarray*}
\xi = \sqrt{\omega_L z},\;\; \xi' = \sqrt{\omega_L z'},
\end{eqnarray*}
Eq.(\ref{HUEN}) becomes,
\begin{eqnarray}
\frac{\partial}{\partial \tau}g(z,\tau|z') = \left\{z\frac{\partial^2}{\partial z^2}+ \left(1+\sigma  -z\right)\frac{\partial}{\partial z} \right\}g(z,\tau|z'),\label{BRLAG}\\
\tau = \frac{a}{\sigma} t,\;\; \sigma = \frac{1}{2(\zeta-1)},\nonumber
\end{eqnarray}
where Eq.(\ref{BRLAG}) is the standard Laguerre Fokker-Planck equation. Hence, due to the initial condition, the time dependent solution for the density in this section is,
\begin{eqnarray}
{\cal T}(x,t|y) = \frac{\exp \left\{ -\frac{(x^{-\frac{1}{\sigma}}+e^{-\frac{a }{\sigma}t}y^{-\frac{1}{\sigma}})}{\omega_L (1-e^{-\frac{a }{\sigma}t})} \right\}I_{\sigma}\left( \frac{2(xy)^{-\frac{1}{2 \sigma}}e^{-\frac{a }{2 \sigma}t}}{\omega_L (1-e^{-\frac{a }{\sigma}t})} \right)}{\sigma \omega_L x^{\frac{3}{2}+\frac{1}{\sigma}}y^{-\frac{1}{2}} e^{-\frac{a}{2}t} (1-e^{-\frac{a }{\sigma}t})}. \label{BRLAG2}
\end{eqnarray}
To the best of our knowledge, Eq.(\ref{BRLAG2}) is a new result of a specific example of a B-R Fokker-Planck equation. 

Making the connection with the Langevin equation this density is generated from,
\begin{eqnarray*}
\dot{x}(t) =a x(t) - \beta  x^{\zeta}(t)\eta_2(t),
\end{eqnarray*}
since for $a>0$ the deterministic system is divergent, but the density is normalisable, this particular case is an example of multiplicative noise stabilising the system \cite{LAST}.

\subsection{Fisher-Snedecor case}
\begin{figure}
\begin{center}
\includegraphics[width=130mm,height=50mm,angle=0]{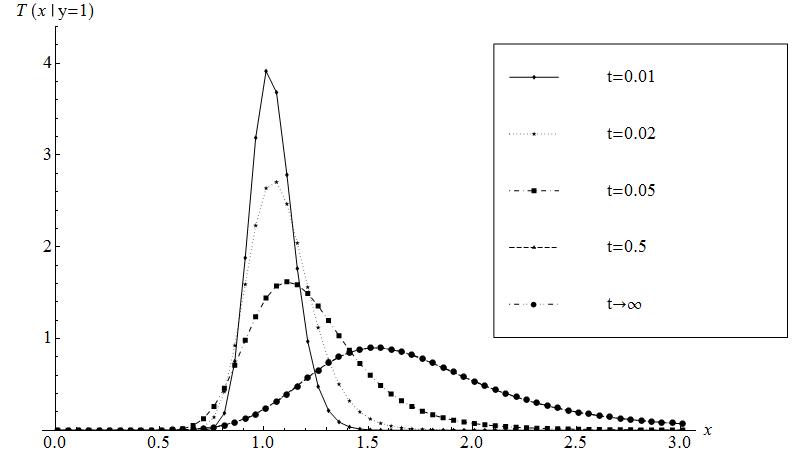}
\caption{Examples of Eq.(\ref{BRFS}) at various times with parameter values $\Omega=1$, $\zeta=3.2$, $a=4.2$, $\alpha=0.9$ and $\beta=0.4$.} \label{fig5}
\end{center}
\end{figure}

Setting $b=0$ the weight function $W(\xi)$ for this case is,
\begin{eqnarray*}
W(\xi)= \xi^{\frac{2\zeta}{\zeta-1} }\left( \xi^2+\omega_F \right)^{\sigma_1-\sigma_2},\\
\omega_F = \frac{\beta^2}{\alpha^2},\;\;
\sigma_1 = \frac{\Omega(\zeta-\frac{1}{2})\alpha^2-a}{2 \Omega (\zeta-1)\alpha^2}-1, \;\; \sigma_2 = \frac{\Omega(\zeta-\frac{1}{2})\alpha^2+a}{2 \Omega (\zeta-1)\alpha^2}  .
\end{eqnarray*}
Applying the following change in variables,
\begin{eqnarray*}
\xi = \sqrt{\omega_F z},\;\; \xi' = \sqrt{\omega_F z'},
\end{eqnarray*}
Eq.(\ref{HUEN}) becomes,
\begin{eqnarray}
\frac{\partial}{\partial \tau}g(z,\tau|z') = \left\{(z^2+z)\frac{\partial^2}{\partial z^2}+ \{2(\sigma_1+1)z +\sigma_1+\sigma_2+1\}\frac{\partial}{\partial z} \right\} g(z,\tau|z'), \nonumber\\ 
\label{BRFISH}\\
 \tau = \frac{2(\zeta-1)a}{\sigma_2-\sigma_1-1} t,\nonumber
\end{eqnarray}
where Eq.(\ref{BRFISH}) is the standard Fisher-Snedecor Fokker-Planck equation. Hence the time dependent solution for the density is,
\begin{eqnarray}
{\cal T}(x,t|y) = \frac{2(\zeta-1)\left( x^{2(1-\zeta)}+\omega_F \right)^{\sigma_1-\sigma_2}}{\omega^{2 \sigma_1+1}_F x^{2 \zeta}} \left\{\sum^{\lfloor -\sigma_1-\frac{1}{2} \rfloor}_{n=0} \frac{(-2n-1 -2\sigma_1)}{n!\Gamma(-n-2\sigma_1)}  \right. \nonumber\\
\times\frac{\Gamma(-n-\sigma_1+\sigma_2) e^{ \frac{2(\zeta-1)an(n+2\sigma_1+1)}{\sigma_2-\sigma_1-1} t } }{\Gamma(1+n+\sigma_1+\sigma_2)} F^{(\sigma_1,\sigma_2)}_n \left(  \frac{x^{2(1-\zeta)}}{\omega_F} \right)F^{(\sigma_1,\sigma_2)}_n \left(  \frac{y^{2(1-\zeta)}}{\omega_F} \right) \nonumber\\
\left. + \int^{\infty}_{0}d \mu \frac{ e^{-\frac{2(\zeta-1)a}{\sigma_2-\sigma_1-1}\left\{ \left(\sigma_1+\frac{1}{2} \right)^2+\mu^2 \right\}t}}{2 \pi \left| \Pi(\mu) \right|^2_{\gamma=\sigma_1}}   \psi_F\left(\mu,\frac{x^{2(1-\zeta)}}{\omega_F} \right) \psi_F\left(\mu,\frac{y^{2(1-\zeta)}}{\omega_F} \right)  \right\}.\label{BRFS}
\end{eqnarray}
As with the Laguerre case, to the best of our knowledge, Eq.(\ref{BRFS}) is a new result of a specific instance of a B-R Fokker-Planck equation. In Fig.\ref{fig5}, we give a specific example of Eq.(\ref{BRFS}) at various times. It is elementary to show that the weight function, which is proportional to the steady state density, peaks at the value $\left( \frac{\Omega \alpha^2-a}{\Omega \zeta \beta^2} \right)^{\frac{1}{2(\zeta-1)}}=1.54$.  

Making the connection with the Langevin equation this density is generated from,
\begin{eqnarray*}
\dot{x}(t) = a x(t) + \alpha x(t) \eta_1(t) - \beta   x^{\zeta}(t)\eta_2(t) ,
\end{eqnarray*}
we see that this example models linear deterministic growth, with linear and quadratic multiplicative stochastic terms. Again, like the Laguerre case, since the deterministic system is divergent, but the density is normalisable, this system provides another example of multiplicative noise stabilising a system \cite{LAST}.

\section{Conclusions and future work}
In this work we have given closed form expressions of sinked densities associated with (at most) quadratic diffusion and linear drift. The eigenfunctions relating to the discrete part of the spectrum are associated variants of classical orthogonal polynomials. We have given a MacRobert's style proof to obtain a new closed form expression for the continuous spectrum normalisation associated with the Romanovski density. This technique is sufficiently generalisable, given one knows enough about the analytic continuation properties of the hypergeometric function under consideration. We then applied these results to obtain the time dependent Fokker-Planck solutions associated with various cases of the B-R Langevin equation.

Given the pervasive nature of Langevin equations (and the densities and Green's functions associated with them) in the physical sciences, we anticipate that these results are a stepping stone to a richer understanding of a variety of processes, both conserved and non-conserved. Specifically, we hope that processes involving mixed spectra eigenfunctions become increasingly  commonplace, as more analytic examples of solution appear which increase our mathematical understanding and our ability to apply such results in novel ways. Paraphrasing the relevant introduction of \cite{BROAD}; in a world of ever increasing computing power, we must never overlook the benefits provided from analytic solutions in terms of special functions. They provide  insight for understanding non-trivial relationships among physical variables with unsurpassed economy of effort, and are an invaluable tool for the validation of more complicated models which require computational treatment.

\ack
The author graciously acknowledges Alexander Kalloniatis for his fruitful discussions and constant encouragement.

\appendix
\section{MacRobert's proof of Eq.(\ref{NEWRESULT})}\label{ROMNORMAPP}
\begin{figure}[htb]
\centering
\includegraphics[width=100mm,height=50mm,angle=0]{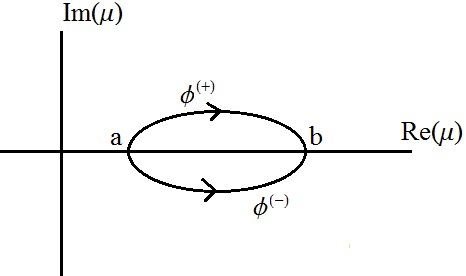}
\caption{Deformed $\mu$ contours $\phi^{(+)}$ and $\phi^{(-)}$ used for the MacRobert's proof.} \label{FIGCONT}
\end{figure}

We begin by conveniently labelling the double integral of Eq.(\ref{NEWRESULT}) by,
\begin{eqnarray*}
{\cal I}_{i,j}(\nu) = \int^{\infty}_{-\infty} dx (x^2+1)^{\gamma}e^{\sigma_2 \arctan (x)} \chi_i(\nu,x) \int^{b}_{a}d\mu \rho_{i,j}(\mu) \chi_{j}(\mu,x),\\
\{a,b\} \in \mathbb{R}_+, \;\;  a < b, \;\; i,j \in \{1,2\}, \;\; \nu  \in \mathbb{R}_+.
\end{eqnarray*}
Focusing on the case $i=j=1$, we apply Eq.(\ref{CHI12}) to split up $\chi_{1}(\mu,x)$ for the region $x > \sqrt{3}$ and deform the $\mu$ integral onto the contours $\phi^{(+)}$ and $\phi^{(-)}$ as shown in Fig.\ref{FIGCONT} to obtain,
\begin{eqnarray}
{\cal I}_{1,1}(\nu) = \int^{\infty}_{-\infty} dx (x^2+1)^{\gamma}e^{\sigma_2 \arctan (x)} \chi_1(\nu,x) \int_{\phi^{(-)}}d\mu \tilde{\Gamma}(\mu) \rho_{1,1}(\mu)\tilde{\chi}(\mu,x)\nonumber\\
+ \int^{\infty}_{-\infty} dx (x^2+1)^{\gamma}e^{\sigma_2 \arctan (x)} \chi_1(\nu,x) \int_{\phi^{(+)}}d\mu \tilde{\Gamma}(-\mu) \rho_{1,1}(\mu)\tilde{\chi}(-\mu,x).\label{INT11}
\end{eqnarray}
Following Chap.14 of \cite{DAVIES} and Chap.7 of \cite{URS}, we may reverse the order of integration as each term in Eq.(\ref{INT11}) falls
off like $x^{-1-Im(\mu)}(1+O(x^{-1})+\dots)$, $Im(\mu) \in \mathbb{R}_+$, on the respective contours $\phi^{(+)}$ and $\phi^{(-)}$, as $x \rightarrow \infty$. Hence ${\cal I}_{1,1}(\nu)$ becomes,
\begin{eqnarray}
{\cal I}_{1,1}(\nu) = \int_{\phi^{(-)}}d\mu \tilde{\Gamma}(\mu) \rho_{1,1}(\mu) \int^{\infty}_{-\infty} dx (x^2+1)^{\gamma}e^{\sigma_2 \arctan (x)} \chi_1(\nu,x)\tilde{\chi}(\mu,x) \nonumber\\
+ \int_{\phi^{(+)}}d\mu \tilde{\Gamma}(-\mu) \rho_{1,1}(\mu) \int^{\infty}_{-\infty} dx (x^2+1)^{\gamma}e^{\sigma_2 \arctan (x)} \chi_1(\nu,x)\tilde{\chi}(-\mu,x). \label{NICEII}
\end{eqnarray}
To proceed we note that (Chap.15.5 of \cite{STEGUN}) $\tilde{\chi}(\pm\mu,x)$, and their complex conjugates, obey the same governing S-L equation as $\chi_1(\mu,x)$ and $\chi_2(\mu,x)$, as they are the corresponding linearly independent solutions in the neighbourhood of the singular point $\infty$. Thus decomposing either of the aforementioned eigenfunctions as,
\begin{eqnarray*}
\{ J_i(\mu,x), \tilde{J}(\mu,x)\} = \sqrt{(x^2+1)^{\gamma+1}e^{\sigma_2 \arctan(x)} }\{\chi_i(\mu,x),\tilde{\chi}(\mu,x)\},
\end{eqnarray*}
the resulting governing equation for the $J$'s is,
\begin{eqnarray}
\left\{ \frac{d^2}{dx^2}+ \left( \frac{\mu^2}{x^2+1} + \frac{x^2 - 4 \gamma \sigma_2 x + 4 \gamma^2 - \sigma^2_2-3}{4(x^2+1)^2}\right)  \right\}\{ J_i(\mu,x), \tilde{J}(\mu,x)\} =0. \nonumber\\
\label{DEFINEJ}
\end{eqnarray}
We now recast Eq.(\ref{NICEII}) in terms of the $J$'s. Through considering two copies of Eq.(\ref{DEFINEJ}), one for $J_1(\nu,x)$ and one for $\tilde{J}(\pm \mu,x)$, we multiply the equation for $J_1(\nu,x)$ by $\tilde{J}(\pm \mu,x)$, and vice versa. Subtracting the two expressions awards us with, 
\begin{eqnarray}
\frac{J_1(\nu,x)\tilde{J}(\pm \mu,x)}{x^2+1}
= \frac{  \tilde{J}(\pm\mu,x)   \frac{d^2}{dx^2} J_1(\nu,x) -    J_1( \nu,x)   \frac{d^2}{dx^2}\tilde{J}(\pm \mu,x) }{\left( \mu^2 - \nu^2 \right) }. \label{unamed}
\end{eqnarray}
Integrating Eq.(\ref{unamed}) over all of $x$, and applying integration by parts we obtain,
\begin{eqnarray*}
 \int^{\infty}_{-\infty}dx \frac{J_1(\nu,x)\tilde{J}(\pm \mu,x)}{x^2+1}= \int^{\infty}_{-\infty}dx (x^2+1 )^{\gamma}e^{\sigma_2 \arctan(x)} \chi_1(\nu,x)\tilde{\chi}(\pm \mu,x) \\
= \left[\frac{\tilde{J}(\pm \mu,x) \frac{d}{dx}J_1(\nu,x)  -  J_1(\nu,x) \frac{d}{dx}\tilde{J}(\pm \mu,x)}{ \mu^2 - \nu^2 } \right]^{x \rightarrow \infty}_{x \rightarrow- \infty}. 
\end{eqnarray*}
Using Eq.(\ref{CHI12}) the asymptotic forms of the desired limits are given by, 
\begin{eqnarray*}
\lim_{x \rightarrow \pm \infty} \tilde{J}(\mu,x)& \sim &{\cal F}_{\pm} (\mu) x^{\frac{1}{2}-i \mu} + O \left(x^{-\frac{1}{2}-i \mu}\right),\\
\lim_{x \rightarrow \pm \infty} \frac{d}{dx}\tilde{J}(\mu,x)& \sim& \pm \left(\frac{1}{2}-i\mu \right){\cal F}_{\pm}(\mu) x^{-\frac{1}{2}-i \mu} + O \left(x^{-\frac{3}{2}-i \mu}\right),\\
\lim_{x \rightarrow \pm \infty} J_1(\nu,x) &\sim&  \tilde{\Gamma}(\nu) {\cal F}_{\pm}(\nu) x^{\frac{1}{2}-i \nu}+ \tilde{\Gamma}(-\nu){\cal F}_{\pm}(-\nu)x^{\frac{1}{2}+i \nu} + O \left( x^{-\frac{1}{2}\pm i \nu} \right), \\
\lim_{x \rightarrow \pm \infty} \frac{d}{dx}J_1(\nu,x) &\sim &\pm \left(\frac{1}{2}-i\nu \right) \tilde{\Gamma}(\nu) {\cal F}_{\pm}(\nu) x^{-\frac{1}{2}-i \nu} \\
 &&\pm \left(\frac{1}{2}+i\nu \right) \tilde{\Gamma}(-\nu) {\cal F}_{\pm}(-\nu) x^{-\frac{1}{2}+i \nu}+ O \left(x^{-\frac{3}{2}\pm i \nu}\right),
\end{eqnarray*}
where,
\begin{eqnarray*}
{\cal F}_{\pm}(\mu) = 2^{\gamma+\frac{1}{2}+ i\mu}  e^{\pm \frac{\pi}{4}\left\{ \sigma_2+2\mu -i (2\gamma+1) \right\}}.
\end{eqnarray*}
Using the above asymptotic forms, Eq.(\ref{NICEII}) becomes,
\begin{eqnarray}
{\cal I}_{1,1}(\nu) =\lim_{z \rightarrow \infty} \int_{\phi^{(-)}}d\mu \rho_{1,1}(\mu){\cal K}(\mu,\nu)\left( \frac{\sin(\mu+\nu)z + i \cos(\mu+\nu)z}{\mu+\nu} \right)\nonumber\\
+\lim_{z \rightarrow \infty}\int_{\phi^{(+)}}d\mu \rho_{1,1}(\mu){\cal K}(-\mu,-\nu)\left( \frac{\sin(\mu+\nu)z - i \cos(\mu+\nu)z}{\mu+\nu} \right)\nonumber\\
+\lim_{z \rightarrow \infty}\int_{\phi^{(-)}}d\mu \rho_{1,1}(\mu){\cal K}(\mu,-\nu)\left( \frac{\sin(\mu-\nu)z + i \cos(\mu-\nu)z}{\mu-\nu} \right)\nonumber\\
+\lim_{z \rightarrow \infty}\int_{\phi^{(+)}}d\mu \rho_{1,1}(\mu){\cal K}(-\mu,\nu)\left( \frac{\sin(\mu-\nu)z - i \cos(\mu-\nu)z}{\mu-\nu} \right),
\label{bigone}
\end{eqnarray}
where $z = \log_e x$ and,
\begin{eqnarray*}
{\cal K}(\mu,\nu) = 2^{2 \gamma+2+i(\mu+\nu)}\tilde{\Gamma}(\mu)\tilde{\Gamma}(\nu) \cosh \frac{\pi}{2}\left( \sigma_2-i(2\gamma+1)+\mu+\nu \right).
\end{eqnarray*}
In the following we consider the \textit{Dirichlet integral} expressions from Chap.1 of \cite{MACROBERT} and Chap.3 of \cite{DAVIES}:
\begin{eqnarray*}
\lim_{z\rightarrow \infty}\int^{\beta}_{-\alpha}d\xi M(\xi) \cos(\xi z) = 0, \;\;\lim_{z\rightarrow \infty} \int^{\beta}_{-\alpha}d\xi M(\xi) \sin(\xi z) = 0,\\
\lim_{z\rightarrow \infty}\int^{\beta}_{-\alpha}d\xi M(\xi) \frac{\cos(\xi z)}{\xi} = 0, \\
\lim_{z\rightarrow \infty} \int^{\beta}_{-\alpha}d\xi M(\xi) \frac{\sin(\xi z)}{\xi} = \frac{\pi}{2} \{M(0+)+M(0-)\},
\end{eqnarray*}
where $\alpha, \beta \in \mathbb{R}_+$ and the analytic function $M(\xi)$ obeys \textit{Dirichlet's conditions} on the interval $(-\alpha,\beta)$\footnote{Dirichlet's conditions for function $M(\xi)$ on the interval $(-\alpha,\beta)$ entail: (I) $M(\xi)$ contains only a finite number of discontinuities on the interval, (II) $M(\xi)$ contains a finite number of turning points on the interval.}. Thus we deform the contours $\phi^{(+)}$ and $\phi^{(-)}$ back to the real line segment $(a,b)$, and let $a \rightarrow 0$ and $b \rightarrow \infty$. Assuming that the function $\rho_{1,1}(\mu)$ obeys Dirichlet's conditions, we immediately obtain the following expression for ${\cal I}_{1,1}(\nu)$,
\begin{eqnarray*}
{\cal I}_{1,1}(\nu) =  \pi \rho_{1,1}(\nu) \left\{ {\cal K}(\nu,-\nu) +{\cal K}(-\nu,\nu) \right\},
\end{eqnarray*}
which is the required form given in Eq.(\ref{NEWRESULT}). The expression for ${\cal I}_{2,2}(\nu)$ is simply the complex conjugate of the case just considered. The remaining cases can be verified in an equivalent method considered in this Appendix.

\end{document}